\documentclass[twocolumn]{article}

\usepackage[leading=normal,charwidths=tight,mathdisplays=normal]{savetrees}
\usepackage{authblk}
\usepackage{amsmath}
\usepackage{graphicx}
\usepackage{multirow}
\usepackage[utf8]{inputenc}
\usepackage[english]{babel}
\usepackage[allcolors=blue,colorlinks=true]{hyperref}
\usepackage[T1]{fontenc}
\usepackage{booktabs}
\usepackage{multirow,dcolumn}
\newcolumntype{d}[1]{D{.}{.}{#1}}
\usepackage{amsmath}

\usepackage{xfrac}

\newcommand{\minus}{\scalebox{.4}[.8]{\( - \)}}
\newcommand{\pone}{1}
\newcommand{\mone}{\minus1}
\newcommand{\zero}{0}
\newcommand{\phlf}{\sfrac{1}{2}}
\newcommand{\mhlf}{\minus\sfrac{1}{2}}

\date{}

\author[1]{Paweł Magnuszewski}
\author[2]{Sylwester Arabas}
\affil[1]{Faculty of Electrical Engineering, Automation, Computer Science and Biomedical Engineering, AGH University of Krakow, Poland}
\affil[2]{Faculty of Physics and Applied Computer Science, AGH University of Krakow, Poland}

\usepackage[backend=biber,style=alphabetic]{biblatex}
\renewbibmacro{in:}{}

\title{Path-dependent option pricing with two-dimensional PDE using MPDATA}

\begin{document}
  \twocolumn[
  \begin{@twocolumnfalse}
    \maketitle
   \begin{abstract}
In this paper, we discuss a simple yet robust PDE method for evaluating path-dependent Asian-style options using the~non-oscillatory forward-in-time second-order MPDATA finite-difference scheme.    
The valuation methodology involves casting~the Black-Merton-Scholes equation as a transport problem by first transforming it into a homogeneous advection-diffusion PDE via variable substitution, and then expressing the diffusion term as an advective flux using the pseudo-velocity technique.
As a result, all~terms of the Black-Merton-Sholes equation are consistently represented using a single high-order numerical scheme for the advection operator.
We detail the additional steps required to solve the two-dimensional valuation problem compared to MPDATA valuations of vanilla instruments documented in a prior study.
Using test cases employing fixed-strike instruments, we validate the solutions against Monte Carlo valuations, as well as against an approximate analytical solution in which geometric instead of arithmetic averaging is used.
The~analysis highlights the critical importance of the MPDATA corrective steps that improve the solution over the underlying first-order "upwind" step.
The introduced valuation scheme is robust: conservative, non-oscillatory, and positive-definite; yet lucid: explicit in time, engendering intuitive stability-condition interpretation and inflow/outflow boundary-condition heuristics.
MPDATA is particularly well suited for two-dimensional problems as it is not a dimensionally split scheme.
The documented valuation workflow also constitutes a useful two-dimensional case for~testing advection schemes featuring both Monte Carlo solutions and analytic bounds.
An~implementation of~the introduced valuation workflow, based on~the PyMPDATA package and the Numba Just-In-Time compiler for Python, is provided as free and open source software.
\end{abstract}
\vspace{.5em}
  \end{@twocolumnfalse}
]

  \section{Introduction} 

  \subsection{Nomenclature}

Asian options are financial derivative instruments whose payoffs depend on the average value of the underlying asset's price. 
These were initially referred to as {\em averaging} options \cite{Bergman_1985} before the term {\em Asian} was adopted \cite{Krzyzak_1990}. 
Such instruments belong to the class of path-dependent options (see, e.g., \cite{Wilmott_et_al_1995} for an introductory overview).
They are categorized as {\em exotic} in contrast to {\em vanilla} European or American options, whose values depend solely on the underlying asset's price at the time of exercise.

The distinction between options that can only be exercised at maturity and those that allow early exercise applies to averaging options as well.
Consequently, both European-style Asian options (also called {\em Eurasian}) and American-style Asian options (also known as {\em Amerasian} or {\em Hawaiian}) appear in the nomenclature \cite{BenAmeur_et_al_1999,Hansen_and_Joergensen_2000}. 
Instruments whose payoffs are defined in terms of the ratio between the average price and the spot price are referred to as {\em Australian} options \cite{Ewald_et_al_2013}.

In market practice, averaging is typically based on the discrete arithmetic mean of the underlying asset price.
However, in~theoretical contexts, both discrete and continuous averaging (see \cite{Fusai_2008} for a detailed discussion), as well as geometric means, are considered.

  \subsection{Payoff structure}
We focus on fixed-strike variant of Asian calls (option to buy) and puts (option to sell) for which the payoff at maturity $f(t=T)$ is:
\begin{eqnarray}
   \text{fixed-strike Asian call:\,\,\,} f(S,A,T) = \max(A(T) - K,\text{~} 0)\\ 
   \text{fixed-strike Asian put:\,\,\,} f(S,A,T) = \max(K-A(T),\text{~} 0)
\end{eqnarray}
where $S(t)$ is the price of the underlying asset evolving in time~$t$, \( A(t)\) is its path-dependent function (e.g., an arithmetic mean), and \( K \) is a fixed strike price.

\subsection{Market usage}
Asian options are used in hedging strategies where average prices are more relevant than spot prices, such as in managing currency exposure \cite{Zhang_1998,McDonald_2006}. 
They are common in commodity markets such as oil \cite{Hruska_2015}, where averaging reduces the effects of price volatility and manipulation.
Insurance companies use Asian options through equity-linked annuities \cite{Lewis_2002}.
Companies sometimes also employ Asian options in share buyback programs to minimize the impact of short-term market fluctuations \cite{Exotic_options_trading}

\subsection{Pricing methods and software}
Valuation of Asian options within the framework of the Black-Merton-Scholes \cite{Black_Scholes_1973,Merton_1973} frictionless-market geometric Brownian-motion risk-neutral model was addressed using several computational methods (see \cite{Privault_2020}, for an overview): PDE methods involving augmentation of the state space to two dimensions \cite{Bergman_1985}, convolution method \cite{Carverhill_and_Clewlow_1990}, Monte Carlo \cite{Kemna_Vorst_1990},
one-dimensional PDE (with analytic approximations) \cite{Ingersoll_1987,Kemna_Vorst_1990,Rogers_and_Shi_1995,Vecer_2001} and binomial trees \cite{Chalasani_et_al_1998}.
Formulations of the problem beyond the geometric Brownian motion model \cite[e.g.,][]{March_2010} as well as data-driven methods  \cite[e.g.,][]{Gan_et_al_2020} have also been proposed.

Asian option valuation routines are part of software packages such as Matlab™, Maple™ and the open-source QuantLib \cite{Varma_et_al_2016} package.
QuantLib uses Monte-Carlo and analytical methods.
Matlab™ documentation \cite{MathWorks_2024} covers analytic, Monte-Carlo and binomial tree methods implemented in the Financial Instruments Toolbox™.
Asian option pricing using numerical solutions to one-dimensional PDE are featured in the documentation of Maple™ \cite{Maple_2025}.

\subsection{Scope of this work}

We focus on pricing fixed-strike arithmetic-average Asian options using the two-dimensional PDE cast as homogeneous advection problem, and numerically solved using a finite-difference scheme.
It is a follow-up to \cite{Arabas_Farhat_2020} where this approach was detailed using the Multidimensional Positive-Definite Advection Transport Algorithm (MPDATA, \cite{Smolarkiewicz_1983,Smolarkiewicz_1984}) for vanilla European and American instruments.
Casting the governing PDE as a transport problem simplifies the numerics, as it yields a single-operator framework.
Employing the MPDATA numerics for discretizing the advection operator ensures high-order and sign-preserving solutions (guaranteeing non-negative price which matches the instrument inherent optionality).
The non-oscillatory variant of the algorithm \cite{Grabowski_and_Smolarkiewicz_1990} addresses simultaneously the notorious issues of numerical diffusion and spurious oscillations that hamper the accuracy of finite-difference solutions to PDE valuation problems \cite[e.g.][]{Zvan_et_al_1996,Milev_and_Tagliani_2013,Cen_et_al_2013,Duffy_2022}.
The~eponymous multidimensional character of MPDATA (i.e., that a two-dimensional step is more accurate than a sum of two single-dimensional steps) makes the algorithm particularly suited for the two-dimensional PDE governing the Asian option valuation.

In the following section~\ref{sec:pde}, the governing PDE and numerical solution method using MPDATA is presented.
Section~\ref{sec:ref} covers reference Monte-Carlo and analytic solutions used to validate the MPDATA results.
Section~\ref{sec:test} discusses the test case, implementation and results used to depict the algorithm operation.
Section~\ref{sec:summary} concludes the paper with a brief summary.

\section{Asian option valuation PDE as a transport problem}\label{sec:pde}

\subsection{Governing 2D PDE}

The considered terminal-value problem is governed by the Black-Merton-Scholes PDE
   in an augmented state-space in which the instrument price $f(S,A,t)$ is
   a function of both the price of the underlying $S$ and the path-dependent $A$:
  \begin{equation}\label{eq:asian_pde}
  \frac{\partial f}{\partial t} + rS \frac{\partial f}{\partial S} + \frac{\sigma^2}{2} S^2 \frac{\partial^2 f}{\partial S^2} + v \frac{\partial f}{\partial A} - rf = 0
\end{equation}
where $T$ is the instrument's time-to-maturity and $v=dA/dt$.
Two alternative ways of defining $A$ have been proposed in literature for PDE (\ref{eq:asian_pde}):
\begin{itemize}
  \item \cite{Ingersoll_1987,Kemna_Vorst_1990}: as a running sum from \( t=0 \) to \( t \), normalized by~\( T \): in which case $A$ corresponds to the  average only at \( t=T \);
  \item \cite{Bergman_1985,Barraquand_1996}: as the actual average from \( t=0 \) to current time \( t \).
\end{itemize}
Equivalence of both approaches was discussed in \cite{Zvan_et_al_1996,Meyer2001}.
Noting that the running sum variant of the formulation avoids a singularity at $t=0$ in the $1/t$ term, we use the normalised running sum variant, which for the case of arithmetic averaging yields:
\begin{equation}\label{eq:A}
    A=\frac{1}{T}\int_0^t S(\tau) d\tau \text{,} 
\end{equation}
what implies constant-in-time and constant-in-$A$ coefficient in the $\partial_A f$ term in the governing PDE:
\begin{equation}\label{dAdt}
v = dA / dt = S / T
\end{equation}
After integrating from $t=T$ to $t=0$, the value of the Asian option as a function of the spot price is inferred as $f(S, A=0; t=0)$.

\subsection{Casting as an advection-diffusion problem}

The governing PDE (\ref{eq:asian_pde}) can be transformed into a homogeneous, constant-coefficient advection-diffusion equation using the substitution (introduced in the context of the original Black-Merton-Scholes PDE for $f(S; t)$ partially in \cite{Brennan_and_Schwartz_1978}; in analogous form as here in \cite{Joshi_2008}, eq. 5.69 therein; for discussion of the interpretation of the transformation see \cite{Arabas_Farhat_2020}):
\begin{equation}
\label{eq:changeVariable}
\left\{
\begin{array}{ll}
    x = \ln S \\
    y = A \\
	\Psi = e^{-rt} f(S, A, t) \\
    u = r - \sigma^2 / 2 \\
    v = e^x / T \\
    \nu = -\sigma^2 / 2 
\end{array}    
\right.
\end{equation}
For arithmetic averaging with a running sum, the resultant transport problem is:
\begin{equation}\label{eq:asian_adv}
      \frac{\partial \Psi}{\partial t} + u \frac{\partial \Psi}{\partial x} + v \frac{\partial \Psi}{\partial y}- \nu\frac{\partial^2 \Psi}{\partial x^2} = 0
\end{equation}
We note that $y$ coordinate is not log-transformed since the valuation requires to read values at A=0.

\subsection{Single-operator formulation using pseudo-velocity}
\begin{figure}
      \centering
      \includegraphics[width=.85\linewidth]{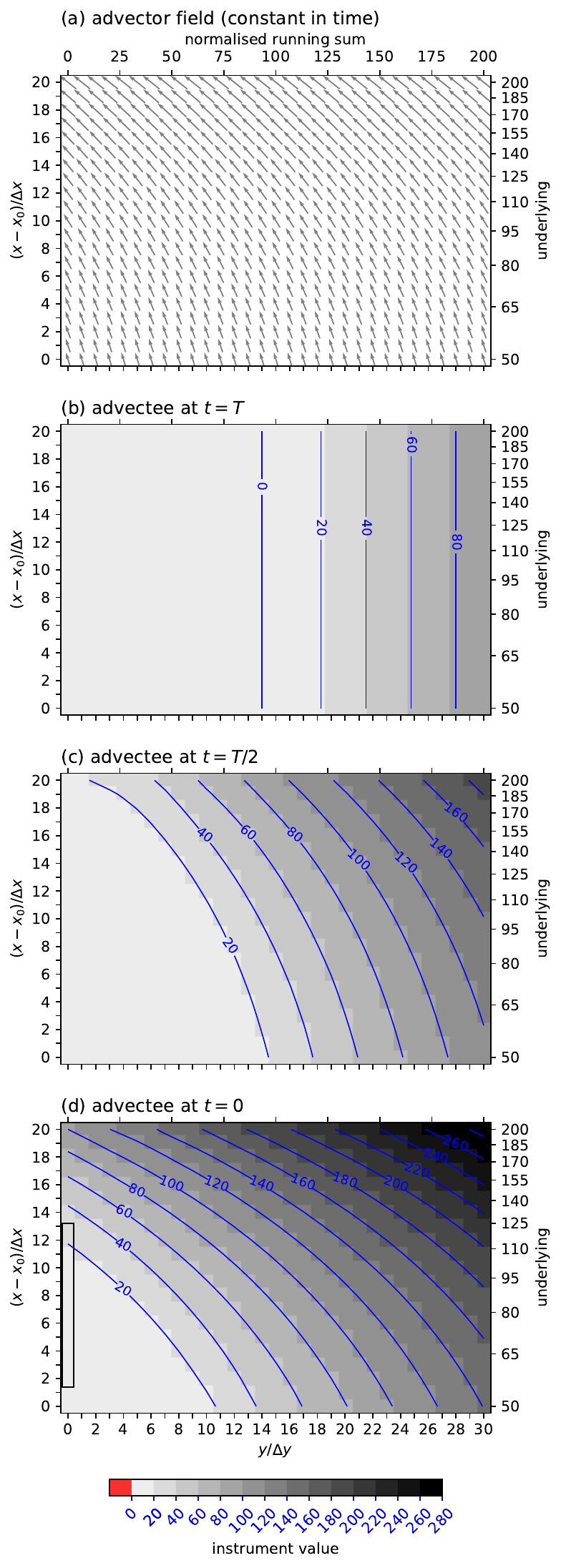}
      \vspace{-1.5em}
      \caption{\label{fig:advectee_over_time}
      Overview of advector and advectee fields for a sample valuation (call payoff, $r=0.08$, $T=1$, $\sigma=0.4$, $21\times 31$ grid).
      (a)~Vector field $\{u,v\}$ (constant-in-time part of $\vec{u}$) plotted with one arrow per grid cell (averaged from the actual Arakawa-C discretisation) and with arrow lengths normalised separately in each direction;
      (b--d)~Scalar field $\Psi$ at three different time steps. Black rectangle in panel (d) indicates location of grid cells plotted in Fig.~\ref{fig:zoom}.
    }  
  \end{figure}

Adapting a general technique described in \cite{Lange_1973,Lange_1978,Smolarkiewicz_and_Clark_1986,Cristiani_2015}, PDE (\ref{eq:asian_adv}) can be transformed into an advection-only problem in ($x$,$y$) coordinates:
\begin{eqnarray}
    \label{eq:asian_advonly}
     \partial_t \Psi +  \nabla \cdot (\vec{u}\, \Psi) = 0  \\
     \vec{u} = \left\{ u - \nu \frac{\partial_x \Psi}{\Psi} ,\, v\right\} \label{eq:pseudo}
\end{eqnarray}
We depict the $(u, v)$ constant-in-time part of the flow field in Figure \ref{fig:advectee_over_time}~(a).
Note that if using geometric mean or an actual average in place of running sum in definition of $A$, the advection-only flux-form formulation (\ref{eq:asian_advonly}) requires an additional right-hand-side term $\Psi \partial_y v$.

  \begin{figure*}[t]
      \centering
      \includegraphics[width=.95\linewidth]{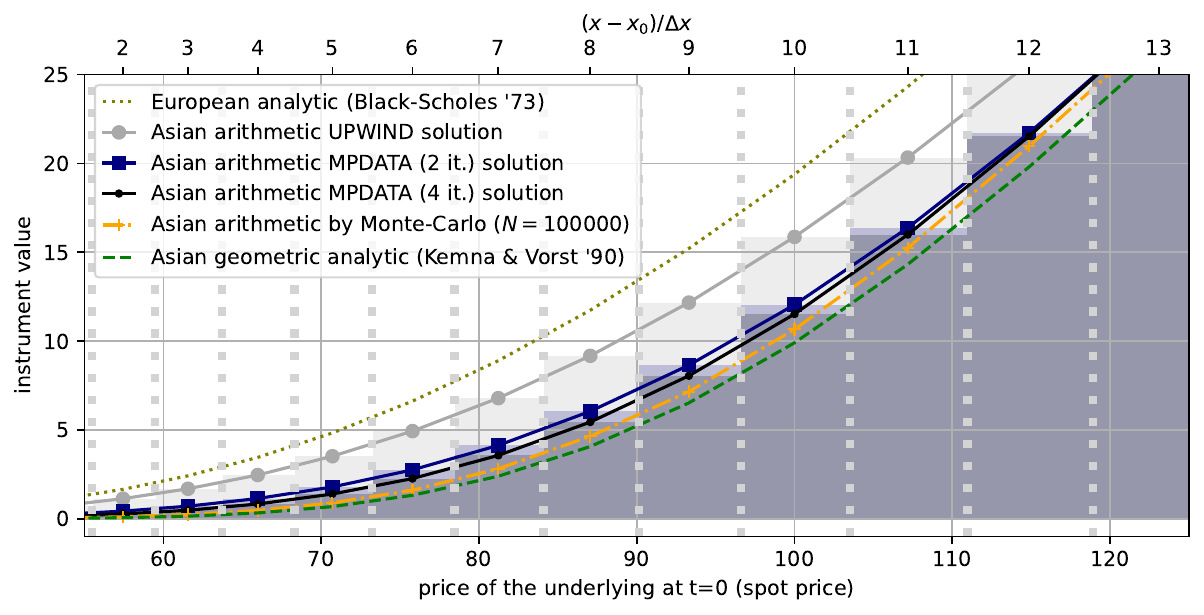}
      \caption{Depiction of values at $t=0$ along $y=0$ transect of the sample valuation domain as in panel (d) in Fig.~\ref{fig:zoom} (see caption for instrument parameters). UPWIND and MPDATA (2 and 4 iterations) results plotted with bins corresponding to grid cell layout and solid lines connecting cell centres. Monte-Carlo solution plotted with dash-dot lines. Analytic solutions for an European option and for geometric-mean Asian option plotted with dotted and dashed lines, respectively.}
      \label{fig:zoom}
  \end{figure*}

\subsection{Iterative finite-difference solution using MPDATA}

The MPDATA scheme, originally developed for atmospheric fluid dynamics \cite{Smolarkiewicz_1983,Smolarkiewicz_1984}, provides a high-order solution to multi-dimensional advection equations of the form (\ref{eq:asian_advonly}).
For~a~review of the algorithm variants and applications, see e.g. \cite{Smolarkiewicz_and_Margolin_1998,Smolarkiewicz_2006,Jaruga_et_al_2015}.

The finite-difference formulation of MPDATA is based on the so-called upstream/upwind/donor-cell  discretisation on a staggered Arakawa-C grid, herein referred to as UPWIND:
\begin{equation}\label{eq:upwind}
  \begin{split}
    \Psi_{[i,j]}^{[n+1]} \!= \Psi_{[i,j]}^{[n]}\!-\!\!\sum\limits_{d=0}^{N-1}\!\left(\!
      F\!\left[\Psi_{[i,j]}^{[n]}, \Psi^{[n]}_{[i,j]+\pi_{\pone,\zero}^{d}}\!, C^{[d]}_{[i,j]+\pi_{\phlf,\zero}^{d}}\right]
      \right.\\
      \left.-
      F\!\left[\Psi_{[i,j]+\pi_{\mone,\zero}^{d}}^{[n]}\!, \Psi^{[n]}_{[i,j]}, C^{[d]}_{[i,j]+\pi_{\mhlf,\zero}^{d}}\right]
    \right)\!\!\!\!
  \end{split}

\end{equation}
with (adopting the notation from \cite{Arabas_et_al_2014}) $\pi^{d}_{a,b}$ indicating cyclic permutation of an order $d$ of an index set $\{a, b\}$:
\begin{equation}
    \sum\limits_{d=0}^{1} \Psi_{[i,j]+\pi_{\pone,\zero}^{d}} \equiv \Psi_{[i+1,j]} + \Psi_{[i,j+1]} 

\end{equation}
where $i,j$ are spatial discretisation indices, $n$ denotes temporal discretisation index, $d$ iterates over $N$ dimensions, $\vec{C}$ is the Courant number vector field discretised at grid cell boundaries denoted here with fractional indices, and $F$ is a flux function defined as:
\begin{equation}
        \begin{split}
    F(\Psi_\text{L}, \Psi_\text{R}, C) = {\rm max}(C,0) \cdot \Psi_\text{L} + {\rm min}(C,0) \cdot \Psi_\text{R}
    \end{split}

\end{equation}
For the present problem, the Courant vector field is:
\begin{equation}
    \vec{C}=\left\{\frac{\Delta t}{\Delta x}\left(u-\nu\frac{\partial_x \Psi}{\Psi}\right), \frac{\Delta t}{\Delta y} v\right\}
\end{equation}
with the x component discretised as:
\begin{equation}
    C^{[0]}_{[i+\phlf,j]} = \frac{\Delta t}{\Delta x} \left(u - \nu\frac{2}{\Delta x} A_{[i,j]}^{[0]}(\Psi)\right)
\end{equation}
with:
\begin{equation}
            A^{[d]}_{[i,j]}(\Psi) = 
      \frac{
        \Psi_{[i,j]+\pi^d_{\pone,\zero}} \!- \Psi_{[i,j]}
      }{
        \Psi_{[i,j]+\pi^d_{\pone,\zero}} \!+ \Psi_{[i,j]}
      }

\end{equation}
(with $A$ set to zero if the terms in denominator sum up to zero).

The UPWIND scheme is conservative and sign-preserving, yet only first-order in time and space, and incurs significant numerical diffusion. Comparing the Taylor expansion of~eq.~(\ref{eq:upwind}) finite-difference approximation with the intended problem~(\ref{eq:asian_advonly}), allows to explain quantify the numerical diffusion, for the leading order term of the difference has a form of a Fickian diffusion \cite{Roberts_and_Weiss_1966}.
In \cite{Smolarkiewicz_1983}, it was proposed to employ the pseudo-velocity technique -- as in eq.~(\ref{eq:pseudo}) -- to remove the numerical diffusion from the solution by UPWIND-integrating backward-in-time with an antidiffusive Courant number analytically derived using the modified equation analysis yielding:
\begin{equation}\label{eq:antidiff}
      \begin{split}
      C'^{[d]}_{[i,j]+\pi^d_{\phlf,\zero}} \!\!\!=\! 
        \left| C^{[d]}_{[i,j] + \pi^d_{\phlf,\zero}} \right| 
        \!\cdot\! \left[ 1 - \left| C^{[d]}_{[i,j]+\pi_{\phlf,\zero}^d} \right| \right] 
        \!\!\cdot\! A^{[d]}_{[i,j]}(\Psi) \\
        - \sum\limits_{q=0, q \ne d}^{N} C^{[d]}_{[i,j]+\pi_{\phlf,\zero}^d} 
        \!\!\cdot \overline{C}^{[q]}_{[i,j]+\pi_{\phlf,\zero}^d} \!\!\cdot B^{[d]}_{[i,j]}(\Psi)
    \end{split}

\end{equation}
where 
  \begin{equation}      
    \begin{split} 
\overline{C}^{[q]}_{[i,j]+\pi^d_{\phlf,\zero}} = \frac{1}{4} \cdot \left(
          C^{[q]}_{[i,j]+\pi^d_{\pone,\phlf}} \!\!+ 
          C^{[q]}_{[i,j]+\pi^d_{\zero,\phlf}} 
        \right.+\\ \left.
          C^{[q]}_{[i,j]+\pi^d_{\pone,\mhlf}} \!\!+ 
          C^{[q]}_{[i,j]+\pi^d_{\zero,\mhlf}}
      \right)
\end{split}
  \end{equation}
  and
  \begin{alignat}{2}
        B^{[d]}_{[i,j]} &= \frac{1}{2}\frac{
      \Psi_{[i,j]+\pi^d_{\pone,\pone}} \!\!+ 
      \Psi_{[i,j]+\pi^d_{\zero,\pone}} \!\!- 
      \Psi_{[i,j]+\pi^d_{\pone,\mone}} \!\!- 
      \Psi_{[i,j]+\pi^d_{\zero,\mone}}
    }{
      \Psi_{[i,j]+\pi^d_{\pone,\pone}} \!\!+ 
      \Psi_{[i,j]+\pi^d_{\zero,\pone}} \!\!+ 
      \Psi_{[i,j]+\pi^d_{\pone,\mone}} \!\!+ 
      \Psi_{[i,j]+\pi^d_{\zero,\mone}}
    }

  \end{alignat}
(with $B$ set to zero if the terms in denominator sum up to zero).\\
Noteworthy, as evident from the $q\ne d$ summation in~eq.~(\ref{eq:antidiff}) and the stencil extents, an MPDATA pass in multiple dimensions is not merely a composition of one-dimensional passes - it~offers enhanced accuracy with respect to dimensionally split composition of one-dimensional passes.

The corrective step can be employed iteratively, with the first corrective iteration addressing the numerical diffusion incurred in the UPWIND integration using the physical/financial velocity field~(\ref{eq:pseudo}).
Subsequent iterations employ the same scheme for correcting the numerical diffusion incurred in the UPWIND steps employed in corrective iterations.
The basic variant of MPDATA yields second-order accuracy in time and space with magnitude of the error diminishing with the number of corrective iterations (see \cite[][Figs 2 and 3]{Arabas_Farhat_2020} for depiction of the spatial and temporal second-order convergence in the case of Black-Merton-Scholes equation integration).

The scheme inherits the key properties of UPWIND, namely: conservativeness and positive-definiteness (non-negativity of option price in this case).
Introduction of the corrective iterations reduces numerical diffusion and leads to second-order convergence in space and time (theoretically the latter for time independent and nondivergentflow field, see discussion in \cite[][sec.~3]{Smolarkiewicz_1984}).
Higher-order convergence can be obtained by substituting the basic antidiffusive Courant number formulation with one analytically derived from higher-order modified equation analysis \cite{Waruszewski_et_al_2018}.
The algorithm has several extensions (for an overview, see \cite{Jaruga_et_al_2015}), among which the non-oscillatory variant \cite{Grabowski_and_Smolarkiewicz_1990}, derived from flux-corrected-transport methods and ensuring monotonicity of the solutions, was used herein.

\subsection{Stability criteria}

For advective transport, MPDATA inherits the stability criterion of the UPWIND scheme, i.e., that the Courant number components are less than unity (and less than half for divergent flow fields).
The stability criterion changes upon incorporation of the pseudo-velocity to represent the diffusive fluxes.
The resultant stability criterion is $(2 \nu \Delta t)/\Delta x^2\le\frac{1}{2}$
(see discussion of eq. 27 in \cite{Smolarkiewicz_and_Clark_1986}).

\subsection{Terminal and boundary conditions}

The terminal condition for the pricing PDE is the discounted payoff $\Psi$ at $t=T$:
\begin{equation}\label{eq:terminal_condition}
    \Psi(S, A, T) = e^{-rT} f(S, A, T)
\end{equation}

As discussed in~\cite{Kemna_Vorst_1990,Arabas_Farhat_2020}, casting the pricing problem as a transport equation offers an intuitive inflow/outflow interpretation of the boundary conditions.
From Fig.~\ref{fig:advectee_over_time}~(a), it can be seen that the boundary condition specification for $\Psi$ at maximum value of $x$ (top edge of the plotted domain) and $y=0$ (left edge of the plotted domain) are irrelevant, for the vector field directs the flow outwards from the domain.
For simplicity and in line with the payoff structure, in the present study (as well as in \cite{Arabas_Farhat_2020}), all inflow boundary conditions are set to spatial extrapolation of $\Psi$ and constant extension of the Courant number vector field.
These conditions also apply to corrective iterations in which the vector field is the antidiffusive Courant number.

For discussions of other formulations of the boundary conditions for the 2D Asian-option valuation PDE, see \cite[][sec.~2]{Kemna_Vorst_1990}, \cite[][sec.~3.3]{Barraquand_1996}, \cite[][sec.3]{Meyer2001}, \cite[][sec.~2]{Cen_et_al_2013} and \cite[][sec.~18.7]{Duffy_2022}.

\section{Reference solutions}\label{sec:ref}
To validate the finite-difference solutions, we refer herein to two alternative valuation methods.

\begin{table*}[t]
\label{table:bp_table}
    \centering
    \small
    \renewcommand{\arraystretch}{1.2}
    
\begin{tabular}{ccr|d{2.3}d{2.3}d{2.3}d{2.3}d{2.3}|d{2.3}d{2.3}d{2.3}d{2.3}d{2.3}}
& & & \multicolumn{5}{l|}{\textbf{Call Option}} & \multicolumn{5}{l}{\textbf{Put Option}} \\
$\sigma$ & $T$ & $K$ & 
 \multicolumn{1}{c}{\rotatebox[origin=l]{90}{\cite{Barraquand_1996}}} &
 \multicolumn{1}{c}{\rotatebox[origin=l]{90}{UPWIND}} & 
 \multicolumn{1}{c}{\rotatebox[origin=l]{90}{\bf MPDATA (2 it.)}} & 
 \multicolumn{1}{c}{\rotatebox[origin=l]{90}{MC $N=10000$}} &
 \multicolumn{1}{c|}{\rotatebox[origin=l]{90}{MC $N=100000$}} &
 \multicolumn{1}{c}{\rotatebox[origin=l]{90}{\cite{Barraquand_1996}}} &
 \multicolumn{1}{c}{\rotatebox[origin=l]{90}{UPWIND}} & 
 \multicolumn{1}{c}{\rotatebox[origin=l]{90}{\bf MPDATA (2 it.)}} &
 \multicolumn{1}{c}{\rotatebox[origin=l]{90}{MC $N=10000$}} &
 \multicolumn{1}{c}{\rotatebox[origin=l]{90}{MC $N=100000$}} \\
\midrule
\multirow{3}{*}{0.2} & \multirow{3}{*}{6} & 100 & 4.55 & 7.12 & 4.77 & 4.50 & 4.47 & 2.10 & 4.61 & 2.39 & 2.09 & 2.09 \\
& & 105 & 2.24 & 4.80 & 2.65 & 2.21 & 2.18 & 4.55 & 7.03 & 4.72 & 4.55 & 4.55 \\
\midrule
\multirow{3}{*}{0.2} & \multirow{3}{*}{12} & 100 & 7.08 & 9.14 & 7.19 & 7.04 & 7.00 & 2.37 & 4.31 & 2.55 & 2.38 & 2.36 \\
& & 105 & 4.54 & 6.71 & 4.76 & 4.51 & 4.47 & 4.36 & 6.38 & 4.49 & 4.36 & 4.36 \\
\midrule
\multirow{3}{*}{0.4} & \multirow{3}{*}{6} & 100 & 7.65 & 9.34 & 7.76 & 7.56 & 7.51 & 5.20 & 6.80 & 5.27 & 5.17 & 5.16 \\
& & 105 & 5.44 & 7.11 & 5.59 & 5.38 & 5.32 & 7.75 & 9.30 & 7.79 & 7.73 & 7.73 \\
\midrule
\multirow{3}{*}{0.4} & \multirow{3}{*}{12} & 100 & 11.2 & 12.5 & 11.3 & 11.1 & 11.0 & 6.46 & 7.68 & 6.53 & 6.46 & 6.46 \\
& & 105 & 8.99 & 10.3 & 9.11 & 8.91 & 8.84 & 8.77 & 9.96 & 8.83 & 8.79 & 8.78 \\

\end{tabular}
    \caption{Comparison of fixed-strike Asian call and put option prices for selected cases given in \cite{Barraquand_1996}.
    Option parameters are defined via volatility $\sigma$, maturity $T$ (presented in months) and strike $K$.
    In all cases, the valuation is for spot price $S_0=100$ and for risk-free interest rate $r=0.1$.
    All prices rounded to 3 significant digits.
    Columns labelled \cite{Barraquand_1996} contain data from Tab.~6 therein.
    UPWIND and MPDATA (2 iterations) columns include finite-difference integration results.
    Columns labelled MC contain Monte-Carlo valuation results for two different path number settings.
    See subsection \ref{sec:discretisation} for other discretisation parameters.
     }
     \vspace{1em}
\end{table*}

\subsection{Analytic solution for geometric averaging}

While the arithmetic average is used for Asian options in market practice \cite{Exotic_options_trading}, an exact closed-form solution exists for both call and put options under geometric averaging~\cite{Kemna_Vorst_1990} in which the $A$ is replaced with $G = \exp\left( \int_0^t \ln(S(\tau)) d\tau \right)$.
The pricing formul\ae~are:
\begin{eqnarray} \label{eq:analytic_solution}
d_1 &=& \Big(\ln{\frac{S}{K}}+ \frac{T}{2}\left(r + \sigma^{2}/6 \right)\Big) / \left(\sigma \sqrt{T/3}\right) \\
d_2 &=& d_1 - \sigma \sqrt{T/3} \\
C_G &=& S e^{-\frac{T}{2}(r+\sigma^2/6)}N(d_1) - K e^{-rT} N(d_2) \\
P_G &=& K e^{-rT} N(-d_2) - S e^{-\frac{T}{2}(r+\sigma^2/6)}N(-d_1)
\end{eqnarray}\vspace{-.5em}\\
where \( N(\cdot) \) is the cumulative standard normal distribution.

Since a geometric average is always less than or equal the arithmetic average, the price of a geometric-average Asian option can be used as a lower bound on the price of an equivalent arithmetic-average Asian option \cite[see discussion in][sect.~4, page 123]{Kemna_Vorst_1990}.

\subsection{Monte-Carlo solution}

For Monte-Carlo pricing, we proceed as described in \cite{Capinski_and_Zastawniak_2012}:
\begin{enumerate}
    \item Simulate a single path of the price of the underlying asset using given initial conditions (spot price $S_0$, volatility $\sigma$, risk-free rate~$r$, and time to expiration $T$) and discretize it into $M$ time steps.
    \item Calculate the payoff of the Asian option based on the average price over the path.
    \item Repeat steps 1 and 2 for $N$ independent paths to generate a distribution of payoffs.
    \item Estimate the option price by averaging the simulated payoffs and discounting back to present value: 
    \begin{equation}        
    C = e^{-rT} \cdot \frac{1}{N} \sum_{j=1}^N \text{Payoff}_j \text{~.}
    \end{equation}
\end{enumerate}
For Monte Carlo simulations, we used $M = 1000$ time steps and the number of paths ranging from $N = 10000$ to $N=100000$. 
\section{Sample results: UPWIND, MPDATA \& Monte-Carlo}\label{sec:test}
\subsection{Definition of the test case}

Figures~\ref{fig:advectee_over_time} and~\ref{fig:zoom}
present the test case explored herein for a single (low resolution) valuation.
Table~\ref{table:bp_table} sums up a series of valuations for different instrument and discretisation parameters.
Specifically, we price a fixed-strike, arithmetic-average Asian call and put options for various combinations of volatility $\sigma$, time-to-maturity~$T$, and strike price~$K$, while keeping the risk-free interest rate constant at $r = 0.1$.
In all valuations, the domain extent for the asset price $S$ is set to range from $S_{min}=50$ to $S_{max}=200$, and the augmented variable $A$ is defined accordingly, with $A_{min}=0$ and $A_{max}=S_{max}=200$.

The results obtained with MPDATA are compared with 
UPWIND solutions, the closed-form solution (Fig.~\ref{fig:zoom} only) and the Monte Carlo pricing method.
In Table~\ref{table:bp_table}, we additionally compare the MPDATA valuations against results obtained using the Forward Shooting Grid (FSG) approach used in \cite{Barraquand_1996}.
FSG  extends the standard binomial pricing model by augmenting the state vector at each node in the lattice tree to capture the option’s path-dependent features.

\subsection{Discretisation parameters}\label{sec:discretisation}

Payoff integration for discretising the terminal condition was done using numerical definite integral (\verb=scipy.integrate.quad=) over each grid cell extents.

The timestep used in the valuations was chosen arbitrarily within the stability limit: $\Delta t=1/500$ (year) was used for the figures, while $\Delta t=1/1760$ was used for the valuations reported in Table~\ref{table:bp_table}. The impact of the number of MPDATA iterations by comparing 2 and 4 iterations in Fig.\ref{fig:advectee_over_time} and presenting UPWIND results in all cases (i.e. 1 iteration, no corrections).

\subsection{Result discussion}

Table \ref{table:bp_table} is organized into 13 columns.
The first three specify the option parameters: volatility $\sigma$, time-to-maturity $T$, and strike price $K$.
The next five columns report prices for the call contract: reference values from \cite{Barraquand_1996}, followed by our UPWIND, MPDATA, and Monte Carlo estimates with $N=10000$ and $N=100000$.
The final five columns replicate this layout for the put contract.

The MPDATA results demonstrate consistent improvement over UPWIND, the latter in most cases deviating from the reference solutions at the first significant digit.
In all presented cases, MPDATA valuations match the reference values from \cite{Barraquand_1996}, comparable in accuracy to the Monte Carlo estimates (for which the convergence is confirmed by lack of significant variation across $N=10000$ and $N=100000$ paths) with a relative tolerance of 20\%.

Presented results  conclude a proof-of-concept stage of the project.
The discretisation parameters were selected arbitrarily, within the stability constraints, to demonstrate the viability of the approach and the radical improvement over UPWIND.

\subsection{PyMPDATA and other MPDATA implementations}

In the present study, we use the Python just-in-time (JIT) compiled open-source implementation of MPDATA -- \verb=PyMPDATA= \cite{Bartman_et_al_2022}.
Presented setup has been incorporated into the package as one of examples in the package documentation since version 1.4.2 (code used to generate all the figures and the table in this paper is persistently archived at \href{https://doi.org/10.5281/zenodo.15543716}{DOI:10.5281/zenodo.15543716}).
JIT compilation and multithreading in PyMPDATA is achieved using Numba \cite{Lam_et_al_2015} which implies that the library code is also valid Python code executable without JIT compilation.
For the valuations presented above, speedups achieved through enabling JIT compilation were approximately $150\times$, $400\times$ and $1000\times$ for UPWIND, 2- and 4-iteration MPDATA, respectively.

For readers interested in implementations of the MPDATA algorithm in other languages, we provide below an overview of key Fortran and C++ versions documented in the literature.

Fortran MPDATA implementations have been developed
as routines embedded in computational fluid dynamics simulations systems such as EULAG \cite{Prusa_et_al_2008,Smolarkiewicz_and_Charbonneau_2013} (atmospheric and astrophysical turbulent [magneto]hydridynamics), ROMS \cite{Thyng_et_al_2021} (ocean modelling) and ECMWF IFS-FVM \cite{Kuehnlein_et_al_2019} (weather forecasting).
A Fortran routine with two-dimensional serial implementation of MPDATA including the non-oscillatory variant \cite{Grabowski_and_Smolarkiewicz_1990} has been open-sourced and released as an electronic supplement to \cite{Grabowski_et_al_2018}.

C++ implementations of MPDATA include the \verb=AtmosFOAM= fork of \verb=OpenFOAM= \cite{Weller_et_al_2022} and the reusable library \verb=libmpdata++= \cite{Jaruga_et_al_2015}, the latter used for European and American option valuation code used in \cite{Arabas_Farhat_2020} (available as electronic supplement to \cite{Arabas_Farhat_2020}).
The third-order option \cite{Waruszewski_et_al_2018} has been implemented in \verb=libmpdata++=.
Both \verb=libmpdata++= and \verb=PyMPDATA= 
feature hybrid shared- (threading) and distributed-memory (MPI) parallelism, both include support for the non-oscillatory algorithm variant.

\section{Summary}\label{sec:summary}
In this work, we have discussed the valuation of path-dependent Asian options using the two-dimensional PDE \cite{Bergman_1985}, casting the governing augmented state-space Black-Merton-Scholes equation as an advection-only transport problem (extending the approach proposed in \cite{Arabas_Farhat_2020} for the basic Black-Merton-Scholes PDE), and solving it numerically using the second-order iterative upwind scheme MPDATA \cite{Smolarkiewicz_1984}.
Numerical integration of all terms in the governing PDE using a single advection operator offers, above all, simplicity (cf. the overview of complexities in numerical PDE approaches to the problem covered in \cite[][chapt.~24]{Duffy_2022}). 
MPDATA guarantees conservative, positive-definite and non-oscillatory solutions addressing notorious challenges in numerical treatment of the Asian option PDE valuation problem \cite{Zvan_et_al_1996,Milev_and_Tagliani_2013,Cen_et_al_2013,Duffy_2022}.
Furthermore, (i)~the algorithm is inherently multidimensional offering improved accuracy over dimension split techniques for Asian option pricing using two-factor PDEs; (ii)~recently developed third-order-accurate variant of MPDATA \cite{Waruszewski_et_al_2018} can be used as a drop-in replacement of the herein employed second-order variant and (iii)~pricing of the American-style instruments using MPDATA has a documented methodology extending the presented approach to a free-boundary problem \cite[][sect.~4]{Arabas_Farhat_2020}.
In addition to contributing to improving the ``finite difference methods gene pool''  \cite{Duffy_2004} for use in computational finance, this work also offers a new two-dimensional test case, with analytic approximations and an established Monte Carlo solution available, to validate numerical advection schemes. 
  
\section*{Acknowledgements}
We acknowledge support from the Polish National Center (grant no. 2020/39/D/ST10/01220) and the AGH Excellence Initiative - Research University (Grant IDUB 9056). The authors thank Ahmad Farhat (who provided the original idea for this study) and Maciej Capiński for discussions  on the preliminary results of this work, as well as Emma Ware for help with library research.

  \printbibliography
\end{document}